\documentclass[12pt,a4paper]{article}

\usepackage{graphicx}
\usepackage{fleqn}
\usepackage{geometry}
\usepackage{amssymb}
\usepackage{amsmath}
\usepackage{caption}
\usepackage{lineno,hyperref}
\usepackage{amsmath,amssymb,amsfonts}
\usepackage{algorithm}
\usepackage{algpseudocode}
\usepackage{textcomp}
\usepackage{float}
\usepackage{subcaption}
\usepackage{makecell}
\newenvironment{keywords}{\small\textbf{Keywords:}\;}{}
\algnewcommand{\LineComment}[1]{\State \(\triangleright\) #1}
\modulolinenumbers[5]
\usepackage{xcolor}

\newtheorem{remark}{Remark}%
\newtheorem{definition}{Definition}%
\newtheorem{statement}{Statement}
\newtheorem{assumption}{Assumption}

\title{Constrained Computational Hybrid Controller for Input Affine Hybrid Dynamical Systems}
\author{
	Ali Taghavian\thanks{Email: taghavianali@gmail.com} \and
	Ali Safi\thanks{Email: ali\_safi@alumni.iust.ac.ir}  \and
	Emaeel Khanmirza\thanks{Email: Khanmirza@iust.ac.ir}  
}

\date{}

\begin{document}
	\maketitle
	\markboth{Espahbodi Nia}{FMTX: An Efficient and Asymptotically Optimal Extension}
\begin{abstract}
	Hybrid dynamical systems are viewed as the most complicated systems with continuous and event-based behaviors. Since traditional controllers cannot handle these systems, some newly-developed controllers have been published in recent decades to deal with them. This paper presents a novel implementable constrained final-state controller based on partitioning the system's state-space, computational simulations, and graph theory. Experimental results and a comparison with Model Predictive Controller on the three tank benchmark and swing-up control of a pendulum show the effectiveness of the proposed Computational Hybrid Controller(CHC).
\end{abstract}
	
	
\section{Introduction}
\begin{table}[htbp]
	\footnotesize		
	\centering
	\caption{Table of Nomenclature}
	\begin{tabular}{p{0.25\linewidth} p{0.6\linewidth}}
		\hline
		\textbf{Symbols} & \textbf{Definition} \\
		\hline
		$x$   & States \\
		$t_{RS}$ & Reachability Sector Applying Duration \\
		$t_{FS}$ & Fine-tuner Section Applying Duration \\
		$k$   & Sampling \\
		$t$   & Time \\
		$q$   & Index of the Rectangular Region \\
		$s$   & Tail Node \\
		$s'$  & Head Node \\
		$J_1$ & Primary Cost Value \\
		$J_2$ & Secondary Cost Value  \\
		$l$   & Center of Elements \\
		$o$   & Operating Node \\
		$Q_1,\:Q_2,\:R$ &  Weighting Matrices for Cost Function  \\
		$T_s$ & Sampling Time \\
		$des$ & Setpoint(Destination)\\
		$\wedge$ & And\\
		$\neg$ & Not\\
		\hline
		\multicolumn{2}{l}{\textbf{Indexes}} \\
		\hline
		$c$   & Continuous \\
		$b$   & Binary \\
		$d$   & Discrete Time \\
		\hline
		\multicolumn{2}{l}{\textbf{Vector \& Matrices}} \\
		\hline
		${{{\mathbf{I}}_{m \times m}}}$   & ($m\times m$)Identity Matrix\\
		${{{\mathbf{0}}_{n \times m}}}$   & ($n \times m$) Matrix of Zeros\\
		${{{\mathbf{1}}_{n \times m}}}$   & ($n\times m$) Matrix of Ones \\
		\hline
		\multicolumn{2}{l}{\textbf{Operators}} \\
		\hline
		$Post(G,s)$   & Head Node Connected to The node s  in The Graph G \\
		$\left\| \cdot \right\|_A^p$ & $L^p$-Norm with Respect to Weight A \\
		$Null(\cdot)$ & Null Space \\
		$\begin{array}{*{20}{c}}
			{\mathop {\min }\limits_x }&{J(x)} \\ 
			{sbj:}&{Ax \leqslant b} 
		\end{array}$ & Minimum Value of Cost Function $J(x)$ for allowable decision variables $x$\\
		$x^+$ &Upcoming Value of x \\
		\hline
	\end{tabular}%
	\label{tab:Nomenclature}%
\end{table}%
\begin{table}[htbp]
	\scriptsize		
	\centering
	\caption{Table of Abbreviation}
	\begin{tabular}{p{0.25\linewidth} p{0.6\linewidth}}
		\hline
		\footnotesize
		\textbf{Abbreviation} & \textbf{Definition} \\
		\hline
		$TG$   & Transition Graph \\
		$RS$   & Reachability Sector \\
		$CHC$   & Computational Hybrid Controller \\
		$DHA$   & Discrete-time Hybrid Automata \\
		$CLF$   & Control Lyapunov Function \\
		$FEM$   & Finite Element Method\\
		$FS$   & Fine-tuner Section \\
		$HA$   & Hybrid Automata \\
		$RS$   & Reachability Sector \\
		$RA$   & Rectangular Automata \\
		$LHA$   & Linear Hybrid Automata \\
		$LP$   & Linear Programming \\
		$MILP$   & Mixed Integer Linear Programming\\
		$MIQP$   & Mixed Integer Quadratic Programming \\
		$MLD$   & Mixed Logical Dynamics \\
		$MMPS$   & Min-Max-Plus-Scale \\
		$MPC$   & Model Predictive Control \\
		$HDS$   & Hybrid Dynamical System \\
		$PWA$   & Piecewise Affine \\
		\hline
	\end{tabular}%
	\label{tab:Abbreviaation}%
\end{table}%

Hybrid systems are formed by the interaction between time-driven and event-driven dynamics. The continuous behavior of the system's dynamics is usually modeled with differential or difference equations, and the discrete one with Finite State Machine. Studying hybrid systems, many dilemmas occur including challenges  in safety, stability, reachability, and controller design.
One of the most interesting challenges is called the \textit{Zeno} phenomenon, which means infinite switching in a limited time. This phenomenon  happens in continuous-time hybrid systems. Many researchers have worked on this issue and how to prevent it from happening because it has destructive effects in most cases. Other challenges include jumping in state variables, switching dynamics, and system constraints.

Since a few decades ago, hybrid systems have been drawing researchers' attention to work on the verification and synthesis problems. Some aspects of hybrid systems like safety, robustness, and reachability properties are studied in the verification process.
On the other hand, the synthesis problem, which is  generally related to modeling and controller design, was developed as soon as the hybrid system concept was introduced.
The most widely used method of modeling \textit{hybrid dynamical systems(HDSs)} is \textit{Hybrid Automata(HA)} \cite{HybridAutomata2003}. Soon
this method was developed and \textit{Discrete-time Hybrid Automata(DHA)} \cite{Torrisi2004} was introduced.
Another popular class of hybrid systems is \textit{Piecewise Affine(PWA)} systems. 
Usually, Lyapunov functions are used to find optimal feedback controllers for PWA systems directly. It is known that Lyapunov functions play an important role in examining the stability and analyzing dynamic systems, and HDSs are no exception. For instance, in \cite{Branicky1998}, some analysis tools were introduced based on Lyapunov functions. Therefore, by designing a controller based on these functions, the stability of systems can be ensured. This type of controller is designed for hybrid systems based on \textit{Control Lyapunov Function(CLF)}, which, if available, can be used to design a stable feedback controller. In general, finding CLFs is very complex for hybrid systems that include both continuous and discrete dynamics \cite{Cairano2014}.
In \cite{Cairano2008} a method for designing this type of controller is presented, which uses two different Lyapunov functions; The first function is a Lyapunov-like function that guarantees finite-time convergence for the discrete state, while the second function guarantees asymptotic stability in the continuous state by local control Lyapunov function. Other Lyapunov-based controllers were developed for different situations \cite{Sanfelice2016,Kersting2017}.

One of the most valuable methods, especially for receding horizon estimations and optimization problems, is \textit{Mixed Logical Dynamics(MLD)}.
While DHA is the most suitable in the modeling phase, MLD is better for solving finite-time optimal control problems \cite{Borrelli2017}. The \textit{Predictive Control} of hybrid systems was investigated in \cite{Borrelli2017}.
\textit{Model Predictive Control(MPC)} is another approach to control PWA systems \cite{Lazar2006}. Besides, with the development of MPC method for MLD systems in \cite{Schutter2004}, it soon became a powerful approach to control HDSs.
Another method to model some of the discrete-event hybrid systems is \textit{Min-Max-Plus-Scaling (MMPS)} expressions \cite{Schutter2001}.

Researches proved that PWA, MLD, DHA, MMPS, and some other methods of modeling hybrid systems are equivalents. For instance, in \cite{DiCairano2010}, a technique was introduced to convert a PWA system to a \textit{Linear Hybrid Automata(LHA)}, or \cite{Bemporad1999} proposed a method to describe DHA as MLD systems.
Another model used for hybrid systems is presented in \cite{Goedel2012} which is a commonly used method for \textit{Impulsive Systems}.
A different method that has a graphical structure, which makes it easier to understand, is \textit{Petri nets}, which is associated with a type of timed network called programmable timed Petri nets \cite{Koutsoukos1999}. They can be used as an alternative to DHAs for two main reasons; the first is expressiveness in the Petri net, and the second is the efficient supervisory controllers that can be designed for discrete-event systems using Petri nets \cite{Moody1997}.
Analyzing the observability and controllability of Petri nets and discussing both timed and untimed Petri nets were investigated in \cite{Silva2011}.
\textit{Supervisory Controllers} were one of the early approaches to control hybrid systems. They are generally intuitive, and they can be mixed with other techniques. For instance, a fuzzy $l$-complete approximation approach was used alongside a supervisory control design to deal with hybrid systems in \cite{Zhang2018}.
In \cite{khanmirza2012}, authors used abstraction of the state-space and then employed graph theory techniques to design a controller in continuous time. At last, they used \textit{Floyd–Warshall} algorithm to find the solution. Furthermore, in another research \cite{Inverted_Girard2012}, a controller for a class of constrained nonlinear systems was designed based on hybridization with triangulation of the state-space. 
It is worth mentioning that several toolboxes are introduced to model hybrid systems. These toolboxes include HYSDEL, HyEQ, Ptolemy, Charon, Modelica, and HyVisual. Also, several toolboxes are introduced to study the safety, reachability, and stability analysis of hybrid systems like CORA, SpaceEx, Hylaa, Julia Reach, Ariadne, DynIbex, Isabelle/HOL, and HyDRA.


In this paper \textit{Computational Hybrid Controller(CHC)} has been presented. This controller is designed to cope with input affine  hybrid systems. The algorithm is based on partitioning the state-space of a dynamical system into regions called \textit{element} which is similar to  \textit{Finite Element Method(FEM)} analyses. 
By assigning an operating node to each element and using them as an initial condition for each element and considering some knowledgeable inputs called \textit{Symbolic Input} a directed graph would be obtained by simulating the system from the initial element to the target element. Analyzing the weighted graph by a graph theory method like \textit{Dijkstra}, the shortest path from each element's operating node toward the final setpoint is found, and the system is controlled. This graph-based controller is called \textit{Reachability Sector}. Since the continuous (and discontinuous) values in an element map to a single point(operating node), there is no guarantee that if the system is within the element but does not run from the operating node, it experiments with the expected trajectory(obtained in RS). So, a secondary controller is defined to drive the system toward the element's operating node. This controller is called \textit{Fine-tuner Section}. Finally, a local stabilizer controller is defined to keep the system on the setpoint.

This paper is organized as follows: The modeling of the hybrid system and allowable dynamics for the CHC method in section 
\ref{modeling}. The controller structure and CHC algorithm in section 
\ref{structure}.
The results of simulations and experiments made on the examples in section
\ref{result}. Finally, some important notes and conclusion in section
\ref{conc}.

\section{System Dynamics and Modelization}\label{modeling}
Almost all controllers work for a specific type of dynamics. There are several methods to model a hybrid dynamical system which are mentioned above. Among all methods, HA is the most complicated modeling method that can model almost all types of hybrid systems. So it is necessary to define the HA system.

\begin{definition}[Hybrid Automata]
	A hybrid automata is defined by nine-tuple sets
	such that:
	\begin{equation}\label{ha}
		H=h(X,Q,U,F,Init,E,G,R,D)
	\end{equation}
	in \eqref{ha} each set represents:
	\begin{itemize}
		\item 
		X: is a finite number of states where $X_c$ denotes the continuous states and $X_b$ denotes the discrete one.
		\item
		$Q=\begin{Bmatrix}
			q_1,q_2,\cdots,q_k
		\end{Bmatrix}$ is a finite set of discrete or binary variables representing different modes.
		\item 
		U: is the set of inputs where $U_c$ is used for continuous inputs and $U_b$ for discrete or binary one. 
		\item 
		$F:X\times U\times Q\rightarrow \mathbb{R}^n $ is the update rule for state variables. n shows number of state variables. Also, $F_c$ represents the update function for continuous states and $F_b$ for discrete or binary one.
		\item 
		$Init\subseteq X$: is the set of initial conditions.
		\item 
		$E\subset Q\times Q$: is a set of possible transitions between different modes.
		\item 
		$G:E\rightarrow 2^{X\times U}$ assigns to each possible mode switching a specific condition which is called guard condition.
		\item 
		$R:E \times X \times U\rightarrow 2^X$ assigns to each guard condition a state reset map.
		\item 
		D(q): Defines the domain of allowable values for states in each location. 
	\end{itemize}
\end{definition}
\begin{assumption}[Existence and Uniqueness]
	The HA must be non-blocking and deterministic. Also, the continuous part of the state update function must be Lipschitz continuous.
\end{assumption}
These conditions must be met to ensure the existence and uniqueness of the simulation
\begin{assumption}\label{assum2}
	the number of system states must be limited, and they should also be bounded.
\end{assumption}
Assumption  
\ref{assum2}
guarantees that the HA state-space can be partitioned into bounded regions. Generally, the shape of the elements is not actually important. However, considering the elements in a rectangular form simplifies the controller design. Rectangular Automata and region graphs are concepts used for reachability analysis
\cite{RA}.
Still, it is needed to modify their definition to apply them in this paper. 

\begin{definition}[Bounded Rectangular Region]
	A \textit{Bounded Rectangular Region} is a subset of an $n$ dimensional space where each state is bounded in the form of open or closed intervals.
\end{definition}

Using the concept of rectangular regions all hybrid automata's locations can be expressed with more limited rectangular regions. 

\begin{definition}[Bounded Rectangular Automata]
	A Bounded \textit{Rectangular Automata}(RA) is a translation of hybrid automata, which is made up of a union of rectangular regions expressed as domain of the location D(q) where a unique state update function is assigned to each rectangular region. Also, all state values inside the $q$th rectangular region are mapped to a single point called \textit{Operating Node}($o^q$) of the rectangular region. RA definition is relatively close to HA.
	\begin{equation}\label{ra}
		RA=h(X,Q,U,F,Init,E,G,R,D,O,L) 
	\end{equation}
	in \eqref{ra} each guard conditon is expressed in form of
	$G(q,q^\prime)=\left\{b<x_j<a\right\}_{j=1}^{m}$ and 
	$R(q^\prime,q)={o^q}$
	where q denotes the index of bounded rectangular region and m denotes the number of states. Moreover, the set L is a set of each element's middle point, which is used for dynamics approximation inside each region. Obviously, a translation of HA to RA must meet the following conditions:
	\begin{enumerate}
		\item 
		The union of RA must cover the whole domain of the original HA, which means $\displaystyle\bigcup_{i\in Q_{RA}}D_{RA}(i)=\displaystyle\bigcup_{i\in Q_{RA}}D_{HA}(i)$
		\item 
		All rectangular regions must be disjointed which means 
		$  \displaystyle\bigcap\limits_{i\in Q}
		D(i)=0$
	\end{enumerate}
\end{definition}
In this paper the continuous dynamic evolution is considered to be in the form of input affine function:
\begin{equation}\label{iaform}
	\dot{x}_c=T^q(x)+G^q(x)u
\end{equation}
in which $T^q$ and $G^q$ are the state-transition and input-output functions at the element $q$, respectively. This assumption will be used 
in section
\ref{fs}. 

\section{Controller Structure}\label{structure}

\begin{figure}
	\centering
	\includegraphics[width=0.4\linewidth]{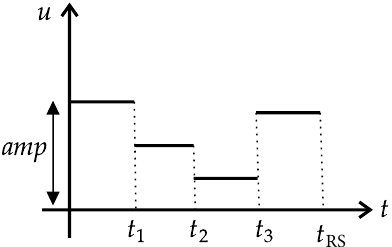}
	\caption{Form of symbolic inputs}
	\label{fig:inputs}
\end{figure}

CHC controller comprises three switching controllers: Reachability Sector, Fine-tuner Section and Stabilizer. The general idea of the controller is based on partitioning the state space of a HDS into a union of elements.
By introducing \textit{Symbolic Inputs} as knowledgeable possible input signals at each location, and executing the RA from all operating nodes as an initial condition, a directed multi-graph from the initial element's operating node to the target element can be generated.This directed multigraph is the main database to control the system named \textit{Transition Graph}(TG). Symbolic Inputs can be considered as signal with any shapes. However, dividing the simulation duration into equal intervals is recommended, and in each interval, a step input is exerted to the system. This recommendation is for making the runtime much more less by discretizing the system. The shape of the input signals is shown in Fig.~\ref{fig:inputs}.
Noticing that by considering the element's operating nodes as initial conditions and defining the reset map to bring the trajectory to the target element's operating node,  the continuous evolution of states is changed  into somehow discrete form at which the system is always mapped into the operating nodes.  Applying graph theory methods to find the shortest path toward the set point like Dijkstra  an analyzed directed  graph will be obtained which shows the trajectory, needed for transition from any initial condition toward the set point. This directed graph is called \textit{Reachabilioty Sector}(RS). This name is adopted from what researchers do to investigate the reachability properties of a HA with some known algorithms like forward and backward reachability tests
\cite{alur2000discrete}. 

\begin{definition}[Transition Graph]
	\textit{Transition Graph}(TG) is a directed multi graph obtained by the simulation of  system for all elements as initial condition for all symbolic inputs. This graph is generated using operating nodes as vertices($V\in O$) and connecting the initial element's operating node as the tail($s\in O$) to the target operating node as the head($s^\prime\in O$) 
	\begin{equation}
		TG=\left<\overrightarrow{s,s^\prime},J_2\right>
	\end{equation}
\end{definition}
\begin{definition}[Reachability Sector]
	\textit{Reachability Sector}(RS) is a directed graph(D) obtained by analyzing the transition graph to find the shortest path between every node to the target set point element using graph theory methods.
	\begin{equation}
		RS=\left<\overrightarrow{s,s^\prime}\right>=Dijsktra(TG,des)
	\end{equation}
\end{definition}
Three steps should be taken to transform the TG to RS:
\begin{enumerate}
	\item 
	Considering at most one connection between two nodes:  
	Since it is possible to have multiple edges from the same tail and head(multigraph) according to Fig.~\ref{fig:ptraject},
	one of these trajectories should be chosen. By defining the \textit{Primary Cost Value}, the nearest target point to the operating node of the target element is chosen as the unique connection between the initial element and the target element.
	\begin{equation}\label{pcost}
		J_1=\left\Vert x-o^q\right\Vert ^{p}_{Q_1} \quad where \quad x\in D(q)
	\end{equation}
	$Q_1$ is the weight matrix with the appropriate dimension, and $p$ is the order of the norm.
	\begin{figure}[ht]
		\centering
		\includegraphics[width=0.4\linewidth]{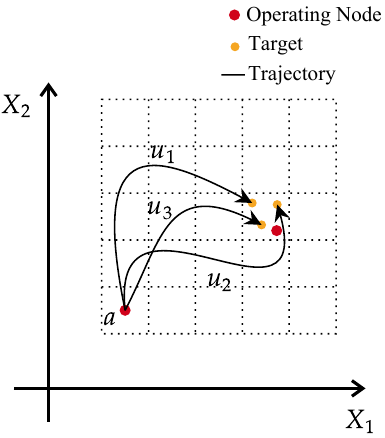}
		\caption{same target for different inputs}
		\label{fig:ptraject}
	\end{figure}
	\item 
	Weighting the edges: The weight of each transition is defined  with \textit{Secondary Cost Value}. Noticing that  the weight of each edge must be defined in a way that avoids unsafe regions. The general form of the cost function can be considered as:
	\begin{equation}\label{scost}
		J_2=\left\Vert o^q-o_{des}\right\Vert ^{p}_{Q_2} +\left\Vert u\right\Vert ^{p}_{R} \quad 
	\end{equation}
	In 
	\eqref{scost},
	$des$ denotes the setpoint that the system is aimed to reach, and $q$ represents the index of the target element. 
	\item 
	Obtaining RS: Analyzing the  weighted Transition Graph by graph theory methods like Dijkstra, Bellman-Ford,etc. The RS will be achieved.
\end{enumerate}

The main issue is: can the continuous state-space inside an element be mapped into one single point and just analyze the graph? Theoretically, it is not justifiable. To resolve this, another controller must be defined to relate the continuous state evolution inside an element to its operating node called Fine-tuner Section(FS). 
Pre-existing methods like the eigenstructure assignment, pole-placement, and robust control approaches are not applicable \cite{khanmirza2012}.
A controller based on Putzer theory was introduced to meet the FS's purpose in \cite{khanmirza2012}, but it was developed only for continuous time problems, and there were no limits on input signals.
In this work,  solution based on an optimization problem for the FS is presented.
\begin{definition}[Fine-tuner Section(FS)]\label{fs}
	Fine-tuner Section is a local controller activated to take the states inside a rectangular region to the vicinity of the element's operating node in a short finite time without violating the element's boundaries.
\end{definition}
Considering the dynamics of the continuous system in each element mentioned in 
\eqref{iaform},
$\dot x_c \in {\mathbb{R}^{n \times 1}}$ can be found by knowing the initial position ($x_0$) in the element when FS is activated.
\begin{equation}\label{initialxdot}
	{\dot x} \mid_{{x_0}} = T^q\left( {{x_0}} \right) + G^q\left( {{x_0}} \right)u \to {{\dot x} \mid_{{x_0}}} = a^q + B^qu
\end{equation}
in which, $a^q \in {\mathbb{R}^{n \times 1}}$, $B^q \in {\mathbb{R}^{n \times m}}$ and $u \in {\mathbb{R}^{m \times 1}}$.
Since the FS operates in a short finite time (${t_{FS}} \in {\mathbb{R}^1}$), in a single time step and also the domain of rectangular regions is small, the next value of the states can be approximated as:
\begin{equation}\label{xplus}
	{\dot x} \mid_{{x_0}}=\frac{x^+-x_0}{t_{FS}}\Rightarrow
	{x^ + } = {x_0} +  {\dot x} \mid_{{x_0}}{t_{FS}}
\end{equation}
The main goal is to reach neighborhood of the operating node, so the cost function of the optimization problem is ${{{\left\| {{o^ q} - {x^ + }} \right\|}^1}}$.
$L^1$-norm is used due to simpler formulation and faster calculation. Therefore, the optimization problem can be written as:
\begin{equation}\label{Min00}
	\begin{array}{*{20}{c}}
		{\mathop {\min }\limits_{u,{t_{FS}}} }&{{{\left\| {{o^q} - \left( {{x_0} + \left( {a^q + B^qu} \right){t_{FS}}} \right)} \right\|}^1}}\\
		{sbj:}&{\begin{array}{*{20}{c}}
				{x_a^q \le {x_0} + \left( {a^q + B^qu} \right){t_{FS}} \le x_b^q}\\
				{{u_{\min }} \le u \le {u_{\max }}}\\
				{t_{FS}^{\min } \le {t_{FS}} \le t_{FS}^{\max }}
		\end{array}}
	\end{array}
\end{equation}
in which ${x_a^q}$ and ${x_b^q}$ are the boundaries of element $q$, ${{u_{\min }}}$ and ${{u_{\max }}}$ are the lower and upper band of input signals, ${t_{FS}^{\min }}$ equals to zero and ${t_{FS}^{\max }}$ is the maximum operating time of the FS. Hence, the control inputs ($u$) and the time needed to reach neighborhood of the operating node ($t_{FS}$) can found by solving the problem \eqref{Min00},. It is obvious that the cost function of the problem \eqref{Min00} is not linear due to the multiplication of $t_{FS}$ in $u$. So by considering $v = u.{t_{FS}}$ as a new variable, we have:
\begin{equation}\label{Min02}
	\begin{array}{*{20}{c}}
		{\mathop {\min }\limits_{v,t} }&{{{\left\| {{o^q} - {x_0} - \left( {a^q.{t_{FS}} + B^qv} \right)} \right\|}^1}} \\ 
		{sbj:}&{\begin{array}{*{20}{c}}
				{x_a^q \leqslant {x_0} + a^q.{t_{FS}} + B^qv \leqslant x_b^q} \\ 
				{{u_{\min }}.{t_{FS}} \leqslant v \leqslant {u_{\max }}.{t_{FS}}} \\ 
				{t_{FS}^{\min } \leqslant {t_{FS}} \leqslant t_{FS}^{\max }} 
		\end{array}} 
	\end{array}
\end{equation}
Problem \eqref{Min02} should be converted to a standard form of convex optimization. Accordingly, a new variable is defined ${y_{\left( {1 + m} \right) \times 1}} = {\left[ {\begin{array}{*{20}{c}}
			{{t_{FS}},}&v^T 
	\end{array}} \right]^T}$. Furthermore by defining ${M_{n \times \left( {m + 1} \right)}} = \left[ {\begin{array}{*{20}{c}}
		{a^q,}&B^q 
\end{array}} \right]$ and ${e_{n \times 1}} = {o^q} - {x_0}$, problem \eqref{Min02} can be rewritten as follow:
\begin{equation}\label{Min04}
	\begin{array}{*{20}{c}}
		{\mathop {\min }\limits_y }&{{{\left\| {e - My} \right\|}^1}} \\ 
		{sbj:}&{\begin{array}{*{20}{c}}
				{a^q.{t_{FS}} + B^qv \leqslant x_b^q - {x_0}} \\ 
				{ - a^q.{t_{FS}} - B^qv \leqslant  - x_a^q + {x_0}} \\ 
				{ - {u_{\max }}.{t_{FS}} + v \leqslant 0} \\ 
				{{u_{\min }}.{t_{FS}} - v \leqslant 0} \\ 
				{{t_{FS}} \leqslant t_{FS}^{\max }} \\ 
				{ - {t_{FS}} \leqslant  - t_{FS}^{\min }} 
		\end{array}} 
	\end{array}
\end{equation}

By converting problem \eqref{Min04} into matrix form, we have:
\begin{equation}\label{Min06}
	\begin{array}{*{20}{c}}
		{\mathop {\min }\limits_y }&{{{\left\| {e - My} \right\|}^1}} \\ 
		{sbj:}&{\begin{array}{*{20}{c}}
				{My \leqslant x_b^q - {x_0}} \\ 
				{ - My \leqslant  - x_a^q + {x_0}} \\ 
				{\left[ {\begin{array}{*{20}{c}}
							{ - {u_{\max }}}&{{{\mathbf{I}}_{m \times m}}} 
					\end{array}} \right]y \leqslant {{\mathbf{0}}_{m \times 1}}} \\ 
				{\left[ {\begin{array}{*{20}{c}}
							{{u_{\min }}}&{ - {{\mathbf{I}}_{m \times m}}} 
					\end{array}} \right]y \leqslant {{\mathbf{0}}_{m \times 1}}} \\ 
				{\left[ {\begin{array}{*{20}{c}}
							1&{{{\mathbf{0}}_{1 \times m}}} 
					\end{array}} \right]y \leqslant t_{FS}^{\max }} \\ 
				{\left[ {\begin{array}{*{20}{c}}
							{ - 1}&{{{\mathbf{0}}_{1 \times m}}} 
					\end{array}} \right]y \leqslant  - t_{FS}^{\min }} 
		\end{array}} 
	\end{array}
\end{equation}
which can be written in a more straightforward form, like in equation \eqref{Min08}.
\begin{equation}\label{Min08}
	\begin{array}{*{20}{c}}
		{\mathop {\min }\limits_y }&{{{\left\| {e - My} \right\|}^1}} \\ 
		{sbj:}&{{A_{ineq1}}y \leqslant {b_{ineq1}}} 
	\end{array}
\end{equation}
where:
\[
{A_{ineq1}} = \left[ {\begin{array}{*{20}{c}}
		{{M_{n \times \left( {m + 1} \right)}}} \\ 
		{ - {M_{n \times \left( {m + 1} \right)}}} \\ 
		{\begin{array}{*{20}{c}}
				{ - {u_{\max }}}&{{{\mathbf{I}}_{m \times m}}} \\ 
				{{u_{\min }}}&{ - {{\mathbf{I}}_{m \times m}}} \\ 
				1&{{{\mathbf{0}}_{1 \times m}}} \\ 
				{ - 1}&{{{\mathbf{0}}_{1 \times m}}} 
		\end{array}} 
\end{array}} \right],{b_{ineq1}} = \left[ {\begin{array}{*{20}{c}}
		{{{\left( {{x^q_b} - {x_0}} \right)}_{n \times 1}}} \\ 
		{{{\left( { - {x^q_a} + {x_0}} \right)}_{n \times 1}}} \\ 
		{{{\mathbf{0}}_{m \times 1}}} \\ 
		{{{\mathbf{0}}_{m \times 1}}} \\ 
		{t_{FS}^{\max }} \\ 
		{ - t_{FS}^{\min }} 
\end{array}} \right]\]

Since the equation \eqref{Min08} is similar to the \textit{Basis Pursuit} problem, it can be converted to linear programming \cite{boyd2004}.
Assuming a new auxiliary variable as $z \in {\mathbb{R}^{n \times 1}}$, Problem \eqref{Min08} is converted to a linear programming(LP) optimization problem as shown in equation \eqref{Min10}
\begin{equation}\label{Min10}
	\begin{array}{*{20}{c}}
		{\mathop {\min }\limits_y }&{{{\mathbf{1}}_{1 \times n}}z} \\ 
		{sbj:}&{\begin{array}{*{20}{c}}
				{{A_{ineq1}}y \leqslant {b_{ineq1}}} \\ 
				{\|{e - My}\| \leqslant z} 
		\end{array}} 
	\end{array}
\end{equation}
Problem \eqref{Min10} should be written in the standard LP form, Thus a new variable is defined as: ${p_{\left( {n + 1 + m} \right) \times 1}} = {\left[ {\begin{array}{*{20}{c}}
			{{z^T},}&{{y^T}} 
	\end{array}} \right]^T}$. 
Consequently, the new problem would be:
\begin{equation}\label{Min12}
	\begin{array}{*{20}{c}}
		{\mathop {\min }\limits_p }&{fp} \\ 
		{sbj:}&{\begin{array}{*{20}{c}}
				{{A_{ineq1}}y \leqslant {b_{ineq1}}} \\ 
				{ - z - My \leqslant  - e} \\ 
				{z + My \leqslant e} 
		\end{array}} 
	\end{array}
\end{equation}
Problem \eqref{Min12} can be written in matrix form as it is seen in \eqref{Min14}.
\begin{equation}\label{Min14}
	\begin{array}{*{20}{c}}
		{\mathop {\min }\limits_p }&{fp} \\ 
		{sbj:}&{\begin{array}{*{20}{c}}
				{{A_{ineq1}}y \leqslant {b_{ineq1}}} \\ 
				{{A_{ineq2}}p \leqslant {b_{ineq2}}} 
		\end{array}} 
	\end{array}
\end{equation}
in which:
\[{A_{ineq2}} = \left[ {\begin{array}{*{20}{c}}
		{ - {{\mathbf{I}}_{n \times n}}}&{ - M} \\ 
		{ - {{\mathbf{I}}_{n \times n}}}&M 
\end{array}} \right],{b_{ineq2}} = \left[ {\begin{array}{*{20}{c}}
		{ - e} \\ 
		e 
\end{array}} \right]\]
To eliminate variable $y$ in equation \eqref{Min14}, we define ${{A'}_{ineq1}} = \left[ {\begin{array}{*{20}{c}}
		{{{\mathbf{0}}_{\left( {2n + 2m + 2} \right) \times n}}},&{{A_{ineq1}}} 
\end{array}} \right]$. Therefore, the final form of the linear programming optimization problem is achieved as \eqref{Min16}
\begin{equation}\label{Min16}
	\begin{array}{*{20}{c}}
		{\mathop {\min }\limits_p }&{fp} \\ 
		{sbj:}&{\begin{array}{*{20}{c}}
				{{A_{ineq}}p \leqslant {b_{ineq}}} 
		\end{array}} 
	\end{array}
\end{equation}	
where:
\[{A_{ineq}} = \left[ {\begin{array}{*{20}{c}}
		{{A_{ineq2}}} \\ 
		{{{A'}_{ineq1}}} 
\end{array}} \right],{b_{ineq}} = \left[ {\begin{array}{*{20}{c}}
		{{b_{ineq2}}} \\ 
		{{b_{ineq1}}} 
\end{array}} \right]\]

Is it possible to reach the operating node from anywhere inside an element? Generally no. the position of the operating node inside an element is crucial to be located so that it can be assured that transition toward it is possible. In traditional linear systems, the controllability condition satisfies our consideration. On the other hand, this is not applicable in complicated hybrid systems with bounds on inputs and states. The following discussion illustrates the importance of locating the position of the operating node inside the element. 

Firstly, It is necessary to analyze the rate at which the state variables evolve. The rate of change in states is called \textit{Flow}. Considering the approximated dynamic of RA expressed in 
\eqref{initialxdot}, 
The effect of state-space flow is considered in constant vector $a$. So, the flow is just affected by the inputs. There are some directions where the flow generated by the input along them is zero, determined by the null space of the matrix $B^T$. If the space is $n$ dimensional, then $n-m$ null space vectors can be found. Where $m$ is the rank of the matrix $B$. These directions are called \textit{Unactuated Directions} which are defined in 
\eqref{key}.
\begin{equation}\label{key}
	\begin{array}{l}
		D = \left\{ {{D^q}} \right\}_{q = 1}^{\# {\rm{of\:elements}}}\\
		{D^q} = Null({B^{{q^T}}}) = \left[ {\begin{array}{*{20}{c}}
				{n_1^q,}&{n_2^q,}& \cdots, &{n_{n - m}^q}
		\end{array}} \right]
	\end{array}
\end{equation}

Since it is impossible to move along these directions by input, the state-space flow must move the trajectory. To be able to move along the unactuated directions, the inner product of state flow and unactuated directions must be positive. Otherwise, The direction must be reversed. It is an example that illustrates  the position of the operating node is important, and the trajectory inside the element can not move in the reverse direction. In \eqref{primarycon} unactuated directions that the transitions along them is possible is found.
\begin{equation}\label{primarycon}
	n_m^q=\left\{\begin{matrix}
		n_m^q \quad n_m^q\cdot a^q \ge0\\
		-n_m^q \quad n_m^q\cdot a^q <0
	\end{matrix}
	\right.
\end{equation}

\begin{figure}[H]
	\centering
	\includegraphics[width=0.2\linewidth]{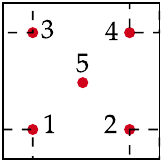}
	\caption{Operating node candidates}
	\label{fig:twodiel}
\end{figure}

The dynamic of the RA must be analyzed to locate each element's operating node. Using discrete form of system dynamics, the operating node is found by minimizing the distance of states after activating the FS controller to the element's operating node. This optimization can be done considering the continuous domain in an element or just choosing some nominees like all corners of the element as operating nodes like what is shown in Figure~\ref{fig:twodiel}. These candidates are known as $o_n^q $, in which $n$ is the number of candidates.
By finding the minimum cost function from the set of costs written in equation
\eqref{oper},
the operating node position inside the element can be found.
\begin{equation}\label{oper}
	\begin{gathered}
		\mathop {\min }\limits_{{o^q}} \{ \begin{array}{*{20}{c}}
			{\mathop {\min }\limits_{u,{t_{FS}}} }&{\sum\limits_{j = 1}^m {\left\| {o_1^q - x{{_j^\prime }^ + }} \right\|_Q^p} , \cdots } 
		\end{array} \hfill \\
		\qquad \qquad\cdots,\begin{array}{*{20}{c}}
			{\mathop {\min }\limits_{u,{t_{FS}}} }&{\sum\limits_{j = 1}^m {\left\| {o_n^q - x{{_j^\prime }^ + }} \right\|_Q^p} } \} 
		\end{array} \hfill \\ 
	\end{gathered}
\end{equation}	
in which $x^\prime$ is a set of $m$ candidate points inside the $q$-th element to investigate the possibility of transition to the operating node from them. 
$x^\prime$ can be considered as only one point in the center of element. $x{_j^\prime }^ +$ can be approximated by equation \eqref{xplus}. Eventually equation 
\eqref{oper}
which is dependent on input and $t_{FS}$ can be solved by the optimization problem \eqref{Min16}.
Finally, the system can be driven toward the setpoint by switching between Reachability Sector and Fine-tuner Section. Another controller can be defined as \textit{stabilizer} controller to stabilize the system and keep it on the setpoint. Depending on the system dynamics, any controller can be used as a stabilizer controller. The Fine-tuner section can not keep the system on the operating node of the element to which the setpoint belongs. So, for the implementation of the CHC, it is essential to use a type of controller with lower computational complexity to stabilize the system on the setpoint. Any type of controller can be used as a stabilizer like PID while the states are inside the element.
Switching between three described controllers the system can be controlled.

Introducing the \textit{Post} operator which gives the head node($s^\prime$) connected to the tail node($s$) of the Graph as:
\begin{equation}
	s^\prime=Post(Graph,s)
\end{equation}
The controller condition to be able to control the system can be expressed in statement 
\ref{reach}
. 
\begin{statement}[Reachability]\label{reach}
	A system can be steered toward the set point if\\ $Post(RS,Init)\ne \emptyset$.
\end{statement}

\begin{remark} 
	Ideally, if the elements size are extremely small, we can assume that by just being in a rectangular region, the states are automatically on the operating node of that element. Therefore, the FS will be disabled automatically by finding $t_{FS}=0$ according to
	\eqref{Min00}.
	So, if $Post(RS,Init)\ne \emptyset$, there exists a path between the initial node to the destination node. This turns out the system can be steered to the set point by applying the exact controller inputs which are found by RS.  
\end{remark}

\begin{remark}
	The size of the elements is a crucial control parameter. It should be chosen  small enough to ensure precise approximations, but not excessively small to prevent  increases in computational burden. In fact, the size of elements depend on several factors including: how fast the dynamic is, input magnitude and its duration, etc. Comparatively to FEM or CFD approaches that there is no "specific" true value for element size, and the solutions are highly dependent on it(it is possible that the solutions become even unreliable for poor element size and quality), this is true for this controller. Since, the RS is  similar to some extent to these methods.
\end{remark}

\begin{figure}
	\centering
	\includegraphics[width=\linewidth]{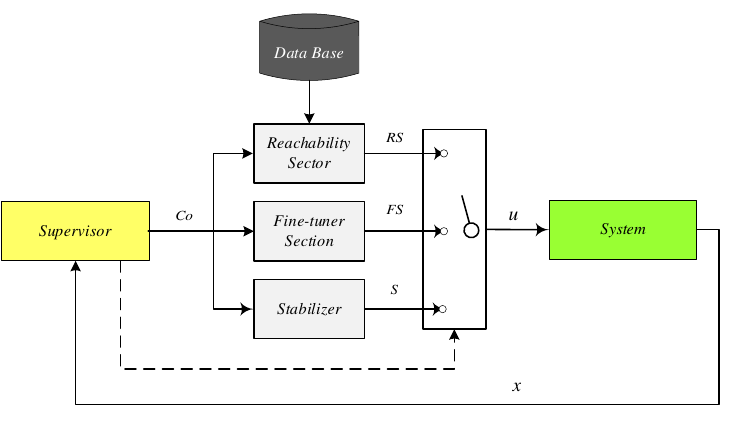}
	\caption{Implementation of CHC}
	\label{fig:CHC}
\end{figure}

The controller structure is shown in Fig.~\ref{fig:CHC}. The Supervisor block is the decision-making unit which activates the appropriate controller in each condition. The flowchart on which the Supervisor block is working is described in 
Algorithm \ref{algo}, in which $Co(k)$ expresses the type of the controller. Also, $\delta_1$ and $\delta_2$ are predefined values showing the allowable distance from the element's operating node and the destination point respectively. 

\begin{algorithm}[!ht]
	\caption{Supervisor Unit Algorithm}\label{algo}
	\begin{algorithmic}
		\State{\textbf{Initialization: $k\leftarrow 2,\:Co(1)\leftarrow FS$}\\
			$\delta_1,\delta_2$}
		\While{True} 
		\If{$\| x(k)-o^q\|\le \delta_1$} 
		\State {$Co(k)\leftarrow RS$} 
		\Else 
		\If{$Co(k-1)=FS$}
		\State{$Co(k)\leftarrow RS$}
		\Else
		\State{$Co(k)\leftarrow FS$}
		\EndIf
		\EndIf
		\If{$\| x(k)-x_{des}\|\le\delta_2$}
		\State{$Co(k)\leftarrow S$}
		\EndIf
		\State{$k\leftarrow k+1$}
		\EndWhile
	\end{algorithmic}
\end{algorithm}

\section{Implementation and Results}\label{result}
CHC controller is easy to implement. This originated from the fact that the CHC controller mainly deals with the Reachability Sector, which is an analyzed graph. So, it is enough to know where the system is, and with no computation, the series of control inputs would be found. To investigate the algorithm's performance , it is implemented on two well-known benchmarks: Three tank and pendulum examples.

\subsection{Pendulum with Input Saturation}
\begin{figure}
	\centering
	\subfloat[Scheme of the pendulum]{\includegraphics[width=0.3\linewidth]{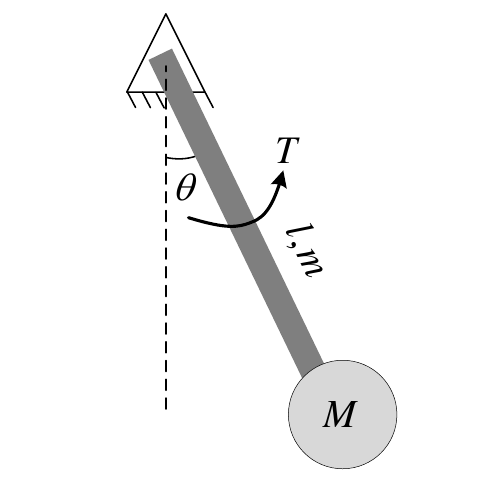}} \qquad
	\subfloat[Experimental setup of the pendulum with input saturation]{\includegraphics[width=0.3\linewidth]{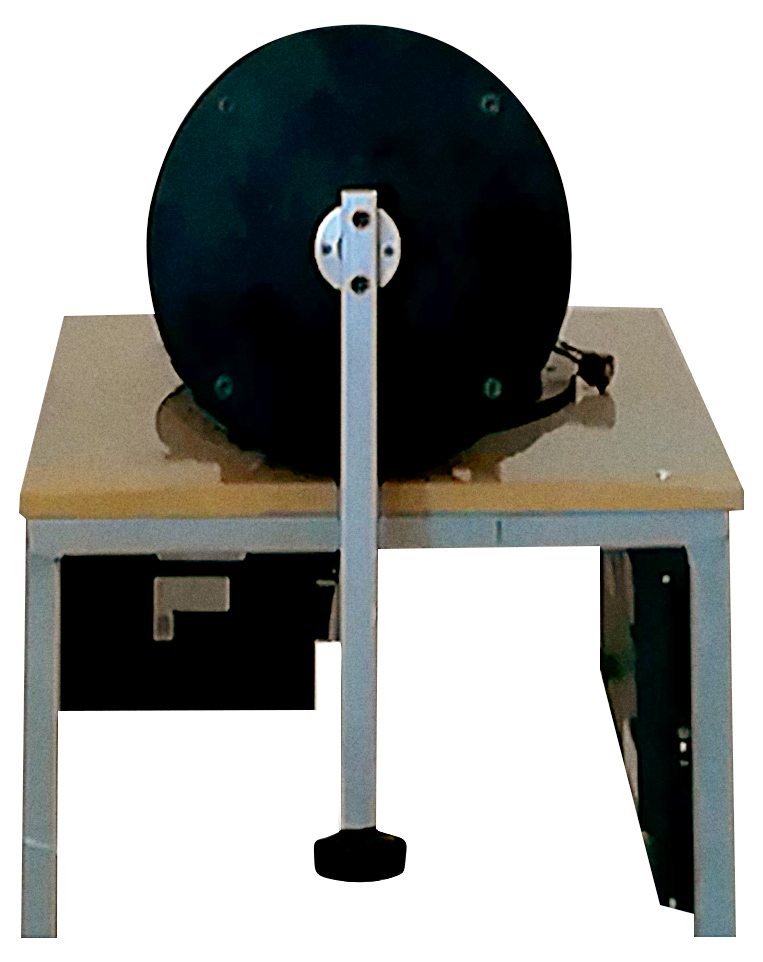}}
	\caption{Scheme and the experimental setup of the pendulum}
	\label{fig:p1}
\end{figure}

The pendulum benchmark is not a hybrid system by itself. However, considering the inputs saturation and modeling it as an RA, makes it a hybrid system. This benchmark has nonlinear dynamics and states change at a high rate. This issue urges the controller to find the inputs fast and accurately. Swing-up problem of the pendulum is much more difficult than only stabilizing it vertically, and most of the controllers are not able to do so by just knowing the setpoint of the problem. This problem is easily solved with the CHC in the existence of input saturation. The problem diagram and the experimental setup are shown in 
\ref{fig:p1}(a) and \ref{fig:p1}(b) respectively. Also, all definitions of the problem are defined in table
\ref{tab:pendulaspe}.

\begin{table}[ht]
	\caption{Parameters used to define the pendulum example}
	\label{tab:pendulaspe}
	\centering
	\begin{tabular}{ccc}
		\hline \textbf{Variable} & \textbf{Description} & \textbf{Value}  \\ 
		\hline $M$ & Pendulum Mass
		& $0.45 \: kg$\\ 
		$m$ & Bar Mass 
		& $0.1\: kg$\\
		$l$ & Bar Length 
		& $0.3\: m$\\
		$c$ & Damping Ratio
		& $0.2\: \frac{T.s}{rad}$\\
		$ g$ & Gravity
		& $9.8\: \frac{N}{kg}$\\
		$ R_\theta$ &  Angle Limit
		& $2\pi\ge \theta\ge 0$\\
		$ R_T$ & Motor Torque
		& $0.9\ge T \ge -0.9$\\
		\hline
	\end{tabular} 
\end{table}

The governing equation of the pendulum is easily written by the angular momentum law about the rotation center:
\begin{equation}\label{dyna}
	T-c\dot{\theta}-(M+\frac{m}{2})glsin(\theta)=(\frac{m}{3}+M)l^2\ddot{\theta}
\end{equation}
changing the variable and considering the equivalent mass as $m^\prime=M+\frac{m}{2}$ and inertia as $I^\prime=(\frac{m}{3}+M)l^2$
The state-space equations become:
\begin{equation}\label{elpen}
	\frac{d}{{dt}}\left[ {\begin{array}{*{20}{c}}
			\theta \\
			{\dot \theta }
	\end{array}} \right] = \left[ {\begin{array}{*{20}{c}}
			{\dot \theta }\\
			{ -\frac{c\dot{\theta}}{I^\prime}- \frac{m^\prime{glsin(\theta) }}{I^\prime} + \frac{T}{I^\prime}}
	\end{array}} \right]
\end{equation}

By partitioning the state-space into rectangular regions and submitting the middle point of each element, the model for each RA becomes:
\begin{equation}\label{lielpen}
	\frac{d}{{dt}}\left[ {\begin{array}{*{20}{c}}
			\theta \\
			{\dot \theta }
	\end{array}} \right] = \left[ {\begin{array}{*{20}{c}}
			{\dot \theta }\\
			{ -\frac{c\dot{\theta_l^q}}{I^\prime}- \frac{m^\prime{glsin(\theta_l^q) }}{I^\prime} + \frac{T}{I^\prime}}
	\end{array}} \right]
\end{equation}

\begin{figure}[H]
	\centering
	\includegraphics[width=0.5\linewidth]{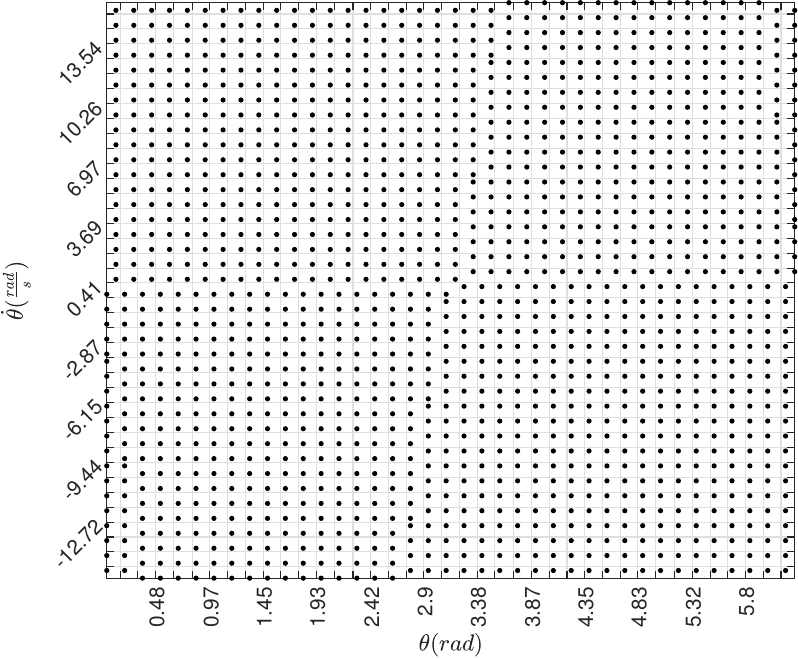}
	\caption{Rectangular regions of inverted pendulum state-space and the operating nodes}
	\label{fig:meshpendula}
\end{figure}

The null space of the matrix $B^T$
is calculated for all locations to be $Null(B^T)=
\begin{bmatrix}
	1, & 0
\end{bmatrix}
$. This implies that inside an element, the input signal cannot move the system in that direction. So, the system must be moved by the state flow vector $a^q$. Applying the relation \eqref{primarycon}, leads to $n_1^q=\begin{bmatrix}
	1, &0 \end{bmatrix}$ if $\dot{\theta}\ge0$ and $n_1^q=\begin{bmatrix}
	-1, &0 \end{bmatrix}$ if $\dot{\theta}<0$. It means if $\dot{\theta}\ge 0$, the operating node must not be chosen such that it is leaned toward the left side of the element since the transition is always in the direction of $n_1^q$ is possible. Five candidate points, according to Fig.~\ref{fig:twodiel} are considered to be the element's operating node. By applying
the equation \eqref{oper} the operating nodes are located and shown in the Fig.~\ref{fig:meshpendula}.

\begin{table}[ht]
	\caption{Simulation Parameters}
	\label{tab:simconpen}
	\centering
	\begin{tabular}{ccc}
		\hline \textbf{Variable} & \textbf{Description} & \textbf{Value}   \\ 
		\hline $Seed$ & \makecell{number of elements\\  in each direction}
		& $\begin{bmatrix}
			40 & 32
		\end{bmatrix}^T$\\ 
		$Q_1$ & \makecell{Primary Cost Function \\ Matrix }  
		& $I_{2\times 2}$\\ 
		$Q_2$ & \makecell{Secondary Cost Function \\ Matrix for states }   
		& $I_{2\times 2}$\\
		$R$ & \makecell{Secondary Cost Function \\ Matrix for inputs}
		& $10^{-6}\times I_{1\times 1}$\\
		$t_{RS}$ & \makecell{Reachability Sector \\ simulation time} 
		& $0.04\: s$\\
		$Init$ &Initial Condition
		& $\begin{bmatrix}
			0 & 0
		\end{bmatrix}^T$\\
		$Des$ & Set Point
		& $\begin{bmatrix}
			\pi & 0
		\end{bmatrix}^T$\\
		\hline
	\end{tabular} 
\end{table}

\begin{figure}[H]
	\centering
	\includegraphics[width=0.5\linewidth]{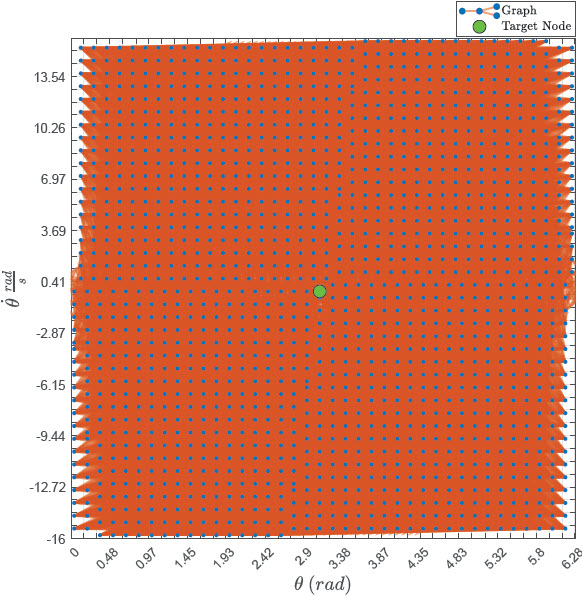}
	\caption{TG for inverted pendulum}
	\label{fig:abspen}
\end{figure}

\begin{figure}[H]
	\centering
	\includegraphics[width=0.5\linewidth]{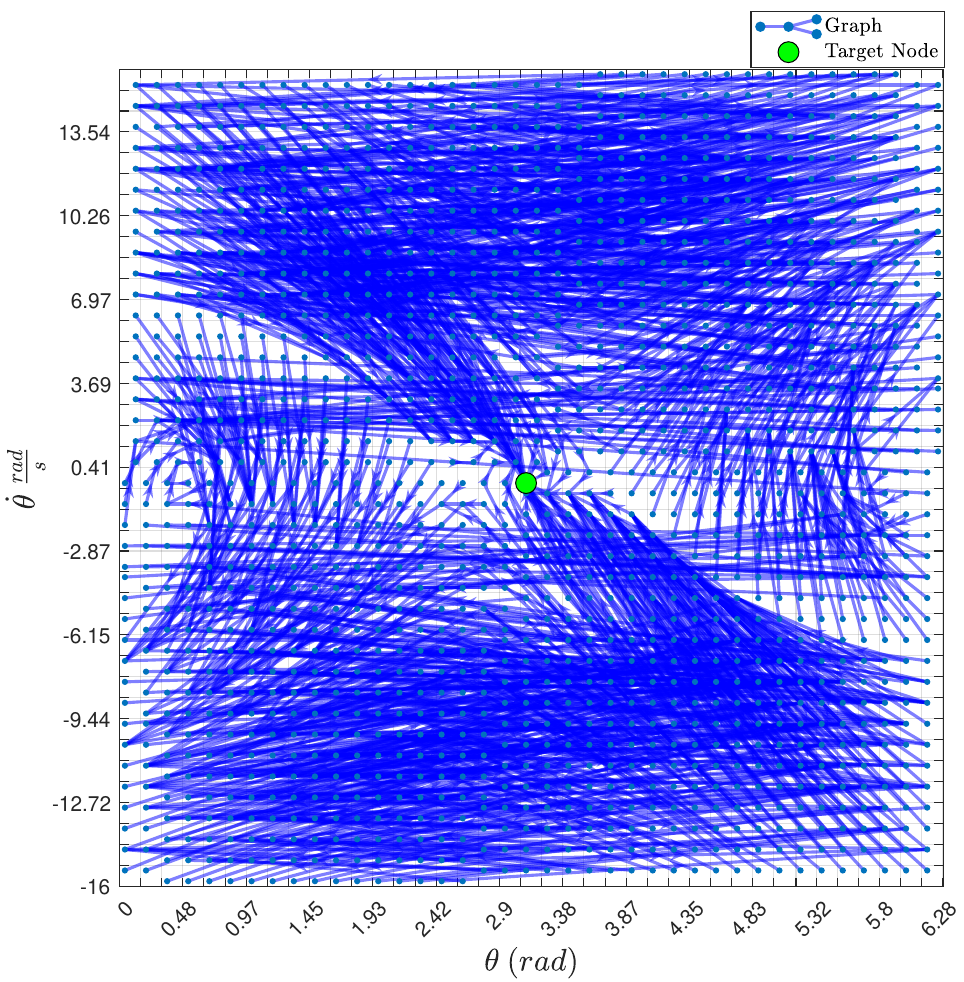}
	\caption{RS for inverted pendulum}
	\label{fig:traabspen}
\end{figure}

To construct the symbolic inputs, the time duration in which the symbolic input is defined is $0.04 \:s$. This time is divided into four equal intervals. at each interval, a different step input is applied to the system. Fig.~\ref{fig:inputs} shows how the symbolic inputs are constructed. In this example amplitude of inputs is considered as 
$amp=
\left[ {\begin{array}{*{20}{c}}
		{ - 0.9,}&{ - 0.75,}&{ - 0.6,}& \ldots, &{0.9}
\end{array}} \right]$.
Simulating the RA with the symbolic inputs, the TG is obtained which is shown in 
\ref{fig:abspen}
. Using the Dijkstra algorithm to find the best route to the setpoint, the Fig.~\ref{fig:traabspen} is obtained. To keep the system on the setpoint when the pendulum is thrown to the vicinity of the setpoint, a local PD stabilizer controller is also used. Considering the system dynamics in equation
\eqref{dyna}, using the controller input written in 
\eqref{controller}
and substituting it in the system dynamics, the coefficients $k_1$ , $k_2$ can be chosen such that stabilize the system.
\begin{equation}\label{controller}
	T=m^\prime glsin(\theta)+c\dot{\theta}-k_1(\theta-\pi)-k_2\dot{\theta}
\end{equation}

\begin{figure}[H]
	\centering
	\includegraphics[width=.55\linewidth]{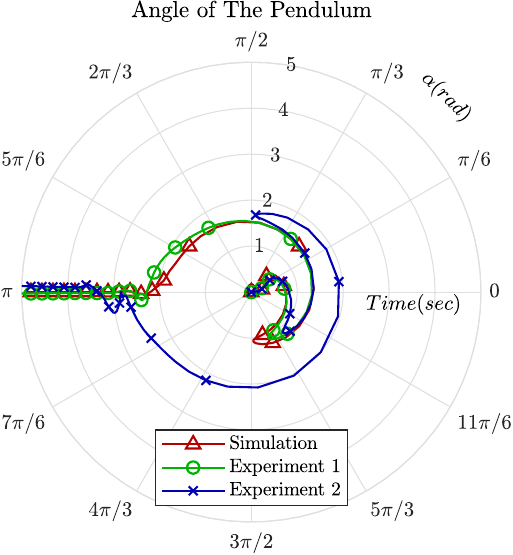}
	\caption{ Evolution of $\theta$}
	\label{fig:intra}
\end{figure}

\begin{figure}[H]
	\centering
	\includegraphics[width=0.6\linewidth]{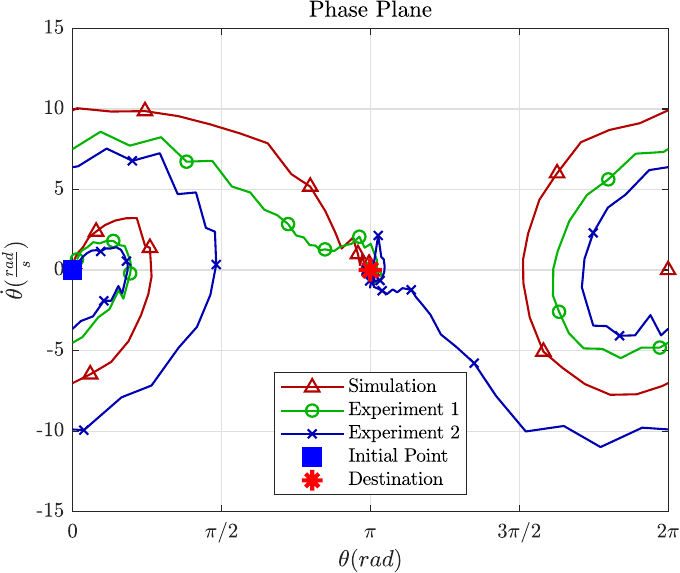}
	\caption{Phase diagram of inverted pendulum}
	\label{fig:pphase}
\end{figure}

\begin{figure}[H]
	\centering
	\includegraphics[width=0.6\linewidth]{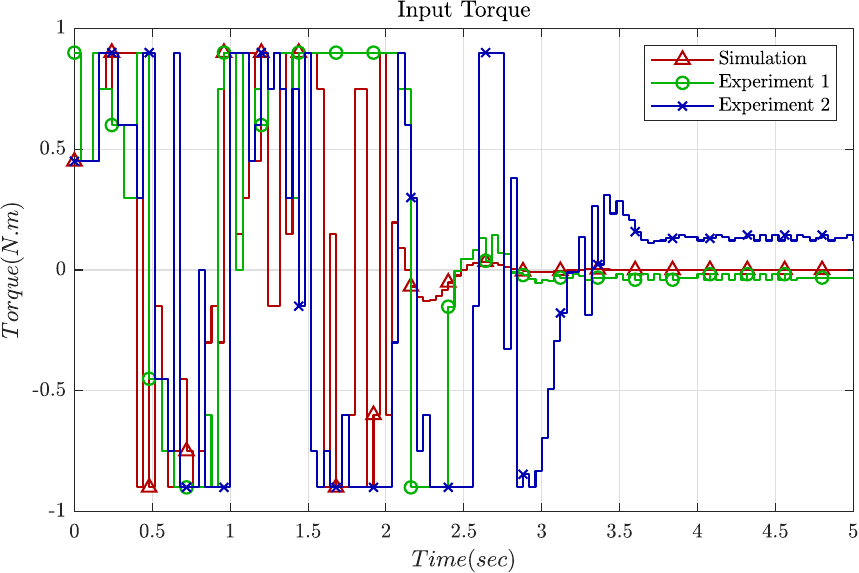}
	\caption{Input of inverted pendulum}
	\label{fig:peneff}
\end{figure}

Finally the algorithm is run to swing up the pendulum. Evolution of $\theta$ to the setpoint is shown in Fig.~\ref{fig:intra} which shows the pendulum is reached the setpoint in 3 swings in the simulation and the first experiment and 4 swing in the second experiment. These fact shows the robustness of the controller. Also, phase diagram is shown in \ref{fig:pphase} and the motor torque figure shown in Fig.~\ref{fig:peneff} indicates that the controller does not violate the saturation limits. The simulation time to reach the Reachability Sector graph with the computer configuration of Intel Core i7-2630QM (2nd Gen) @ 2GHz is approximately 32 minutes. Watch the video of swing-up control of the pendulum with the help of CHC from https://youtu.be/sxKbho5o8mQ .

\subsection{Three Tank}
\begin{figure}
	\centering
	\subfloat[Three tank benchmark scheme]{\includegraphics[width=0.75\linewidth]{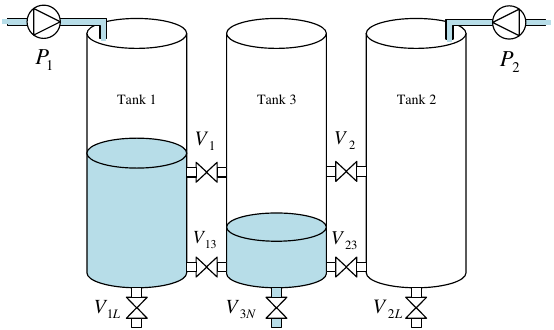}}\\
	\subfloat[Experimental setup of the three tank benchmark]{\includegraphics[width=0.5\linewidth]{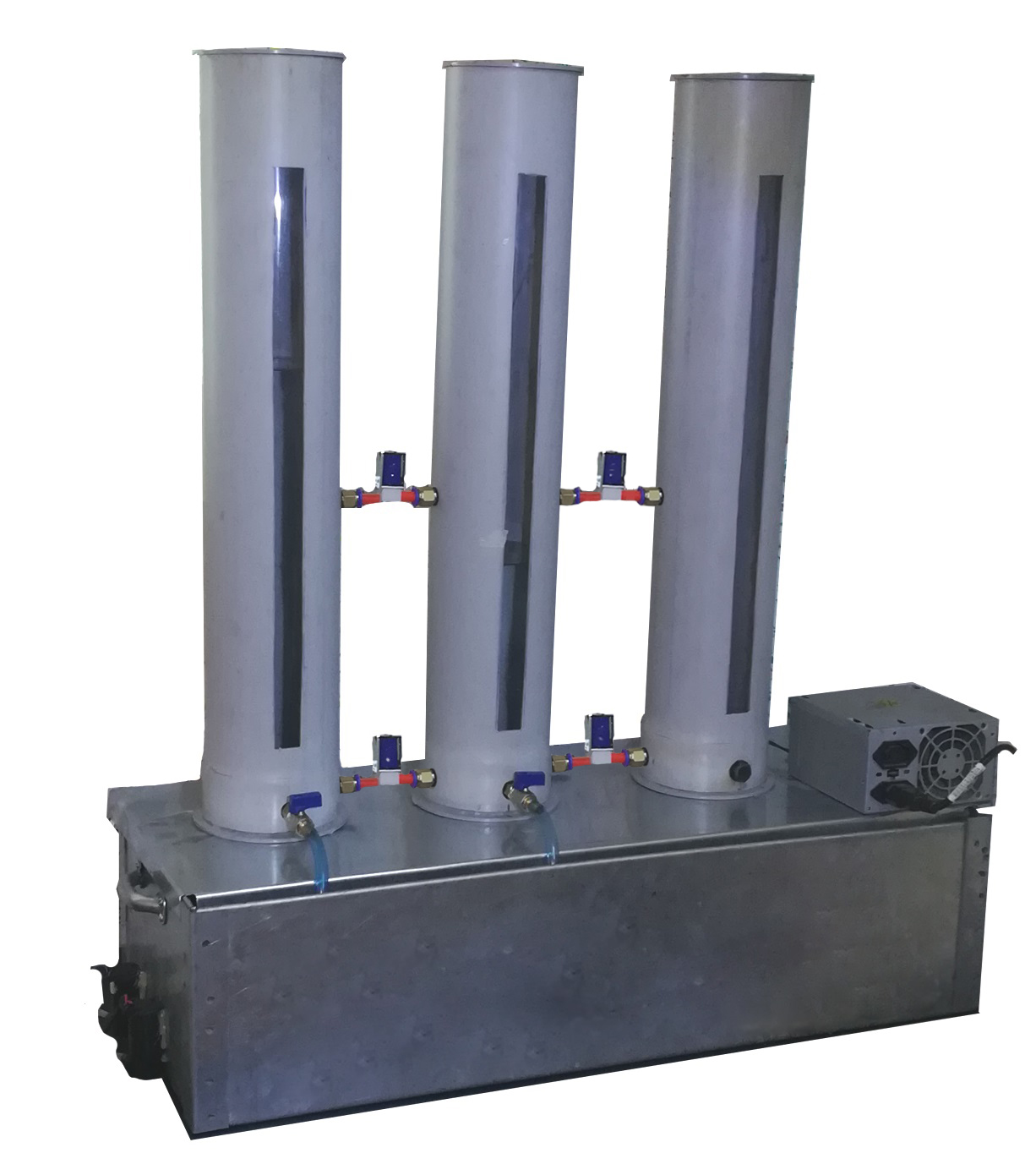}}
	\caption{scheme and the experimental setup of the three tank benchmark}
	\label{fig:threetank}
\end{figure}

Three tank benchmark is one of the most famous examples of hybrid systems. The scheme and the experimental setup of this benchmark is shown in Fig.~\ref{fig:threetank}(a) and \ref{fig:threetank}(b) respectively. Moreover, The dynamics and full explanations of this system can be found in \cite{three}. Also, the PWA model of this system with only two controllable switching valves ($V_{13}\, \& \, V_{23}$) and two continuous flow pumps is specified in equation \eqref{PWA_Threetank}. Note that $V_{13} = u_{b_1}$ and $V_{23}= u_{b_2}$. Also, $u_c$ is the vector of pumps' flows. Furthermore, the specification of the system is expressed in Table~\ref{tab:MotionModels}. Equation \eqref{PWA_Threetank} should be discretized in time with the sampling time of $t_s = 10s$. 
\begin{equation}\label{PWA_Threetank}
	\resizebox{.9\hsize}{!}{$
		\dot x = \left\{ {\begin{array}{*{20}{c}}
				{\left[ {\begin{array}{*{20}{c}}
							{ - {k_1}{A^{ - 1}}}&0&{{k_1}{A^{ - 1}}} \\ 
							0&{ - {k_1}{A^{ - 1}}}&{{k_1}{A^{ - 1}}} \\ 
							{{k_1}{A^{ - 1}}}&{{k_1}{A^{ - 1}}}&{ - (2{k_1} + {k_2}){A^{ - 1}}} 
					\end{array}} \right]x + \left[ {\begin{array}{*{20}{c}}
							{{A^{ - 1}}}&0 \\ 
							0&{{A^{ - 1}}} \\ 
							0&0 
					\end{array}} \right]u_c,if:\,{u_{{b_1}}} \wedge {u_{{b_2}}}} \\ 
				{\left[ {\begin{array}{*{20}{c}}
							{ - {k_1}{A^{ - 1}}}&0&{{k_1}{A^{ - 1}}} \\ 
							0&0&0 \\ 
							{{k_1}{A^{ - 1}}}&0&{ - ({k_1} + {k_2}){A^{ - 1}}} 
					\end{array}} \right]x + \left[ {\begin{array}{*{20}{c}}
							{{A^{ - 1}}}&0 \\ 
							0&{{A^{ - 1}}} \\ 
							0&0 
					\end{array}} \right]u_c,if:\,{u_{{b_1}}} \wedge \neg {u_{{b_2}}}} \\ 
				{\left[ {\begin{array}{*{20}{c}}
							0&0&0 \\ 
							0&{ - {k_1}{A^{ - 1}}}&{{k_1}{A^{ - 1}}} \\ 
							0&{{k_1}{A^{ - 1}}}&{ - ({k_1} + {k_2}){A^{ - 1}}} 
					\end{array}} \right]x + \left[ {\begin{array}{*{20}{c}}
							{{A^{ - 1}}}&0 \\ 
							0&{{A^{ - 1}}} \\ 
							0&0 
					\end{array}} \right]u_c,if:\,\neg {u_{{b_1}}} \wedge {u_{{b_2}}}} \\ 
				{\left[ {\begin{array}{*{20}{c}}
							0&0&0 \\ 
							0&0&0 \\ 
							0&0&{ - {k_2}{A^{ - 1}}} 
					\end{array}} \right]x + \left[ {\begin{array}{*{20}{c}}
							{{A^{ - 1}}}&0 \\ 
							0&{{A^{ - 1}}} \\ 
							0&0 
					\end{array}} \right]u_c,if:\,\neg {u_{{b_1}}} \wedge \neg {u_{{b_2}}}} 
		\end{array}} \right.$}
\end{equation}
\begin{table}[ht]
	\centering
	\footnotesize
	\caption{Model parameters}
	\label{tab:MotionModels}
	\begin{tabular}{ccc}
		\hline \textbf{Variable} & \textbf{Description} & \textbf{Value}   \\ 
		\hline $k_1$ & Connecting Valve Coefficient
		& $3.89\times 10^{-5}\: m^2$\\ 
		$k_2$ & Output Valve Coefficient 
		& $8.65\times 10^{-6}\: m^2$\\
		$A$ & tank Area
		& $0.0123\: m^3$\\
		$h_{max}$ & Max Height
		& $0.66\: m$\\
		$u_{max}$ & max pump flow
		& $2\times 10^{-5}\: m^3$\\
		\hline
	\end{tabular} 
\end{table}

One of the most popular methods used to control the three tank benchmark is MPC \cite{MPC_ThreeTank1,MPC_ThreeTank2,MPC_ThreeTank3}. Therefore, to inspect the effectiveness of the proposed method, both simulation and experimental comparisons between MPC and CHC are provided.
When it comes to hybrid dynamical systems, the Mixed Logic Dynamics(MLD) is one of the best modeling methods, especially for solving optimization problems \cite{Schutter2004}. Accordingly, a model predictive controller can be designed for hybrid systems based on MLD. Thus, HYSDEL 3.0 toolbox \cite{Torrisi2004} is used to extract the MLD model from the PWA model of the three tank example.
Equation \eqref{MPC_ThreeTank} shows the MPC optimization problem, in which the MLD model of the system is considered in constraints.
\begin{equation}\label{MPC_ThreeTank}
	\begin{array}{*{20}{c}}
		{\mathop {\min }\limits_{{U_0}} }&{{J_0}\left( {x\left( 0 \right),{U_0}} \right)} \\ 
		{sbj:}&{\left\{ {\begin{array}{*{20}{c}}
					{{x_{k + 1}} = A{x_k} + B{u_k} + {B_{aux}}{\omega _k} + {B_{aff}}} \\ 
					{{y_k} = C{x_k} + {D_u}{u_k} + {D_{aux}}{\omega _k} + {D_{aff}}} \\ 
					{{E_x}{x_k} + {E_u}{u_k} + {E_{aux}}{\omega _k} \leqslant {E_{aff}}} \\ 
					{{x_N} \in {X_f}} \\ 
					{{x_0} = x\left( 0 \right)} 
			\end{array}} \right.} 
	\end{array}
\end{equation}
in which
$x \in {\mathbb{R}^{{n_{xr}}}} \times {\{ 0,1\} ^{{n_{xb}}}}$,
is a vector of continues and binary state variables,
$u \in {\mathbb{R}^{{n_{ur}}}} \times{\{ 0,1\} ^{{n_{ub}}}}$
is a vector of continues and binary inputs and,
$y \in {\mathbb{R}^{{n_{yr}}}} \times{\{ 0,1\} ^{{n_{yb}}}}$
is a vector of continues and binary outputs. Furthermore,
$w ={[z,\delta]}^T$
is a vector of auxiliary variables, in which
$\delta \in {\{ 0,1\} ^{{n_{d}}}}$
and
$z \in {\mathbb{R}^{{n_{z}}}}$
are continues and binary auxiliary variables correspondingly. All the matrix coefficients are calculated by HYSDEL 3.0 toolbox. In addition, A cost function is minimized to calculate binary(open/close valves) and continues(flow of pumps) input signals in each sampling time. The optimization problem of the MPC becomes a Mixed Integer Linear Programming(MILP), due to existence of binary variables and a $L^1$-norm cost function which is considered as follows:
\begin{equation}
	\begin{gathered}
		{J_0}\left( {x\left( 0 \right),{U_0}} \right) = \left\| {{x_N} - {r_N}} \right\|_P^p +  \hfill \\
		\qquad\sum\limits_k^{N - 1} {\left\| {{x_k} - {r_k}} \right\|_Q^p + \left\| {{u_k}} \right\|_R^p}  + \left\| {{u_b}_k} \right\|_{{R_b}}^p \hfill \\ 
	\end{gathered} 
\end{equation}
where $r$ is the reference point, The Prediction Horizon is $N = 2$ and
\[\begin{gathered}
	P = Q = {{\mathbf{I}}_{3 \times 3}}, \hfill \\
	R = {{\mathbf{I}}_{2 \times 2}},{R_b} = 0.001 \times {{\mathbf{I}}_{2 \times 2}}. \hfill \\ 
\end{gathered} \]

YALMIP toolbox \cite{YALMIP} and CBC solver are used to solve the MILP optimization problem \eqref{MPC_ThreeTank} in each iteration.

To implant the computational hybrid controller in this example, first, we should mesh the state-space. Then by simulating the RA with Symbolic Inputs and analyzing the output graph by the Dijkstra algorithm, the best route from each initial condition toward the setpoint $x_{des}$ can be found. Moreover, CHC's parameters and coefficients are expressed in table \ref{tab:simcon}.
\begin{table}[ht]
	\footnotesize
	\caption{Controller Parameters}
	\label{tab:simcon}
	\centering
	\begin{tabular}{ccc}
		\hline \textbf{Variable} & \textbf{Description} & \textbf{Value}   \\ 
		\hline $Seed$ & \makecell{number of elements\\  in each direction}
		& $\begin{bmatrix}
			10, & 10, & 20,
		\end{bmatrix}^T$\\ 
		$Q_1$ & \makecell{Primary Cost Function \\ Matrix }  
		& $I_{3\times 3}$\\ 
		$Q_2$ & \makecell{Secondary Cost Function \\ Matrix for states }   
		& $I_{3\times 3}$\\
		$R$ & \makecell{Secondary Cost Function \\ Matrix for inputs}
		& $10^{-6}\times I_{2\times 2}$\\
		$t_{RS}$ & \makecell{Reachability Sector \\ simulation time} 
		& $40\: s$\\
		$Init$ &Initial Condition
		& $\begin{bmatrix}
			0, & 0, & 0,
		\end{bmatrix}^T$\\
		$Des$ & Set Point
		& $\begin{bmatrix}
			0.44, & 0.35, & 0.2
		\end{bmatrix}^T$\\
		\hline
	\end{tabular} 
\end{table}

\begin{figure}[!htb]
	\centering
	\includegraphics[width=0.9\linewidth]{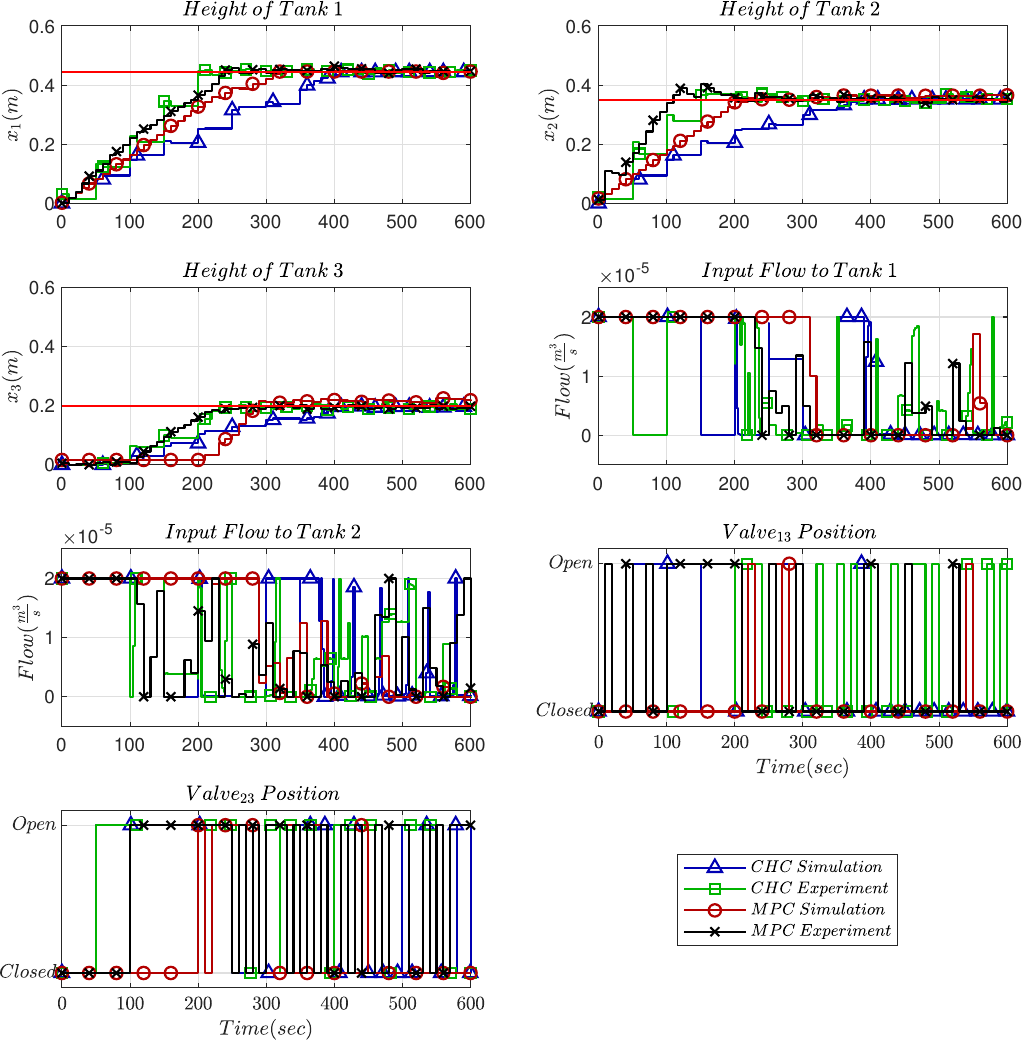}
	\caption{The simulation and experimental results for both CHC and MPC methods on the three tank benchmark}
	\label{fig:ThreeTankAllFigures}
\end{figure}

The simulation and implementations for CHC and MPC are done using the defined specifications. The evolution of the water height in each tank is shown in Fig.~\ref{fig:ThreeTankAllFigures}. This figure also shows that the input flow from the pumps shows that the saturation limit is held.  It should be mentioned that a noise signal with an amplitude of 3 centimeters was added to the simulation while sampling to match the specification of the sensor that we were using to measure the water level.
The simulation time to reach the Reachability Sector graph with the computer configuration of Intel Core i7-2630QM (2nd Gen) @ 2GHz is approximately 11 hours and 23 minutes.

\section{Conclusion}\label{conc}
This paper presents a novel computational controller to reach the desired states of a hybrid dynamical system. This controller operates on input affine systems with bounded states. It has several advantages: the controller easily handles constraints like input saturation and unsafe states; It can be applied to a class of nonlinear systems; The algorithm can be implemented on real-life systems without any time-consuming computation, making it a real-time practical solution. On the other hand, as the number of states grows, the time needed to find the Reachability Sector of the controller increases in the designing phase, which is not desirable. Furthermore, the controller is applied to two different benchmark examples, one of them has nonlinear dynamics, and the other has binary inputs, making it a hybrid system. The simulation and experiment results show the controller's effectiveness in the final-state control problem.

\end{document}